\documentclass[aps,prd,amssymb,eqsecnum,nofootinbib,floatfix, a4paper,twocolumn]{revtex4}

\usepackage{mathrsfs}
\usepackage{latexsym,amsmath,amsfonts,amssymb}
\usepackage{bm}

\usepackage{graphicx}

\allowdisplaybreaks

\newcommand{\be}{\begin{equation}}
\newcommand{\ee}{\end{equation}}
\newcommand{\beq}{\begin{equation}}
\newcommand{\eeq}{\end{equation}}
\newcommand{\bea}{\begin{eqnarray}}
\newcommand{\eea}{\end{eqnarray}}
\newcommand{\eq}{\eqref}
\newcommand{\g}{{\gamma}}

\newcommand{\pinf}{p_{\infty}}

\newcommand{\n}{{\mathbf n}}
\newcommand{\p}{{\mathbf p}}
\newcommand{\vk}{{\mathbf k}}
\newcommand{\bb}{{\mathbf b}}
\newcommand{\ve}{{\mathbf e}}
\newcommand{\bP}{{\mathbf P}}
\newcommand{\cI}{{\cal I}}
\newcommand{\cA}{{\cal A}}

\begin{document}

\title{Radiative contribution to classical gravitational scattering  at the third order in $G$}

\author{Thibault Damour}
\email{damour@ihes.fr}
\affiliation{Institut des Hautes Etudes Scientifiques, 35 route de Chartres, 91440 Bures-sur-Yvette, France}

\date{\today}

\begin{abstract} 
Working within the post-Minkowskian approach to  General Relativity, we prove that the
radiation-reaction to the emission of gravitational waves during the large-impact-parameter scattering of two (classical) point masses
modifies the conservative scattering angle by an additional contribution of order $G^3$ which involves a high-energy (or massless) 
logarithmic divergence of opposite sign to the one contained in the third-post-Minkowskian result of Bern et al.
[Phys. Rev. Lett.  {\bf 122},  201603 (2019)]. The   high-energy limit of the resulting
radiation-reaction-corrected (classical) scattering angle  is finite, and is found to agree with the one following from the (quantum) 
eikonal-phase result of Amati, Ciafaloni and Veneziano [ Nucl. Phys. B {\bf 347}, 550 (1990)].
 \end{abstract}

\maketitle

\section{Introduction}

 Post-Minkowskian (PM) perturbative gravity, i.e., the Poincar\'e-covariant perturbative approach to the classical gravitational 
 dynamics of binary systems,
 was initiated long ago \cite{Bertotti1956,Havas:1962zz,Portilla:1980uz,Westpfahl:1979gu,Bel:1981be,Westpfahl:1985},
 and has been recently revived through the use of new frameworks, notably the effective one-body method 
 \cite{Damour:2016gwp,Damour:2017zjx,Antonelli:2019ytb,Damour:2019lcq},
 and the use of dictionaries relating quantum scattering amplitudes to classical PM dynamics 
 \cite{Damour:2017zjx,Guevara:2017csg,Bjerrum-Bohr:2018xdl,Cheung:2018wkq,KoemansCollado:2019ggb,Kosower:2018adc,Bern:2019nnu,Bern:2019crd,Kalin:2019rwq,Kalin:2019inp,Cristofoli:2019neg,Bjerrum-Bohr:2019kec,Cheung:2020gyp,Parra-Martinez:2020dzs,Kalin:2020mvi,Kalin:2020fhe}.
 Here, we shall focus on the interaction between structureless bodies, see, e.g., Refs.
 \cite{Bini:2017xzy,Vines:2017hyw,Bini:2018ywr,Vines:2018gqi,Guevara:2018wpp,Chung:2018kqs,Guevara:2019fsj,Arkani-Hamed:2019ymq,Siemonsen:2019dsu} for PM-type works on spinning bodies.
 
A recent milestone has been the derivation by Bern et al.  \cite{Bern:2019nnu,Bern:2019crd} of the 3PM-accurate, $O(G^3)$,
conservative gravitational dynamics of binary systems from a two-loop scattering amplitude. 
The 3PM dynamics of Refs. \cite{Bern:2019nnu,Bern:2019crd} 
has been checked at the sixth post-Newtonian (6PN) level \cite{Bini:2019nra,Blumlein:2020znm,Bini:2020wpo}, and
rederived in different ways \cite{Cheung:2020gyp,Kalin:2020fhe}. However, the $O(G^3)$ dynamics of Refs \cite{Bern:2019nnu,Bern:2019crd}
has several puzzling features (see, e.g., the discussion in Ref. \cite{Damour:2019lcq}).  

One of the puzzling features of the 3PM-accurate scattering angle derived in Refs. \cite{Bern:2019nnu,Bern:2019crd} is that it
involves a logarithmically divergent contribution $\propto \ln (s/(m_1 m_2)$ that becomes large both in the 
massless limit ($m_1 \to 0$, $m_2\to 0$) and in the high-energy limit ($ s \to \infty$).
By contrast, previous work by Amati, Ciafaloni and Veneziano (ACV) had extracted
a finite scattering angle from the high-energy (trans-Planckian), two-loop scattering of two massless particles in the (quasi-classical)
eikonal-phase approximation  \cite{Amati:1990xe}. It was argued in Refs. \cite{Bern:2019nnu,Bern:2019crd} that this discontinuity in the
scattering angle was linked to the need of imposing the inequality  $q \sim \frac{\hbar}{b} \ll m_1, m_2$ in order to
extract the classical limit from the quantum scattering amplitude of two massive particles. 
The validity  of the high-energy eikonal ACV scattering angle (as well as
its universality in two-derivative gravity theories) has been recently established  in Refs. \cite{DiVecchia:2019kta,Bern:2020gjj,DiVecchia:2020ymx}. In addition, Ref. \cite{DiVecchia:2020ymx} brought a new light on the puzzling issue of the relation between the
ultra-relativistically-singular classical 3PM dynamics of Refs. \cite{Bern:2019nnu,Bern:2019crd}, and the ultra-relativistic limit of the 
quantum scattering of massive particles (in any two-derivative gravity theory). They presented two different approaches,
and showed that in both cases it is crucial to include radiative effects   for recovering ultrarelativistic finiteness,
and continuity with the ACV eikonal scattering phase \cite{Amati:1990xe}. In particular, in their second approach (involving
the evaluation of a four-point two-loop amplitude in ${\cal N}=8$ supergravity) they emphasized the need to integrate
the graviton contribution over the full soft region, rather than only the potential region.

The results of Ref. \cite{DiVecchia:2020ymx} do not, by themselves, clarify the precise way in which the 
inclusion of {\it classical} gravitational-radiation effects can reconcile the 
(ultra-relativistically-singular) {\it conservative} 3PM dynamics of two {\it classical particles} in General Relativity (GR) with the 
finite eikonal ACV scattering angle. The present work will clarify this issue by proving (by a PM-gravity computation in GR)
 that the classical radiation-reaction to the emission of gravitational radiation during the large-impact-parameter scattering 
 of two classical point masses modifies the conservative scattering angle, $\chi^{\rm cons}$, by an additional contribution, say
 $\chi^{\rm rad}$, which: (i) is of order $G^3$; (ii) cancells the logarithmically divergent high-energy contribution 
 $\propto \ln (s/(m_1 m_2)$ present in $\chi^{\rm cons}$; and (iii)  yields  a finite high-energy 3PM-accurate scattering angle that
 precisely agree with the one following \cite{Ciafaloni:2014esa,Bern:2020gjj} from the ACV eikonal phase \cite{Amati:1990xe}.
  
Amusingly, the resolution offered here of the 3PM high-energy puzzle uses in a crucial way two ideas that were first
discussed many years ago in a completely different context. Indeed, soon after the discovery of the Hulse-Taylor binary pulsar \cite{Hulse:1974eb}
(which brought, after a few years, direct evidence for the reality of gravitational radiation reaction \cite{Taylor:1979zz})
several authors
emphasized the lack of a consistent derivation of radiation-reaction effects in GR \cite{Ehlers:1976ji}. One of the reasons for doubt
was that the then extant (heuristic) derivations of radiation-reaction effects were based on formal {\it post-Newtonian} (PN) expansions involving
infrared ambiguities because of a lack of clear matching between the nearzone gravitational field (``potential modes")
and the wavezone gravitational field (``soft modes").  The answer to this doubt was brought by the use of a
 PM framework in which the gravitational binary dynamics was described by (nonlinearly iterated) {\it retarded-propagator} interactions,
 thereby evacuating the issue of the nearzone/wavezone matching \cite{Westpfahl:1979gu,Bel:1981be,Damour1983}. This led to the derivation of equations of motion
 including (on the same  footing) both conservative and radiation-reaction effects, and leading to observable predictions
 for the binary dynamics that agreed with binary-pulsar observations \cite{Damour:1983tz}. Within this PM framework, it was emphasized in Ref. \cite{Damour:1981bh},
 that the retarded PM equations of motion (which included radiation-reaction effects) implied a loss of the mechanical angular momentum
of the system of order $G^2/c^5$, i.e., at the 2PM level, while they
 implied  a loss of the mechanical energy of the system at the 3PM level (namely $O(G^3/c^5)$).
 Our derivation below of the radiation-reaction correction $\chi^{\rm rad}$ to the scattering angle will be crucially based on 
 a PM computation of the radiative loss of angular momentum at the $O(G^2)$ level. Another crucial ingredient of our computation
 will be to use a general result \cite{Bini:2012ji} linking $\chi^{\rm rad}$ to the radiative losses of angular momentum and energy. 

 We leave to our concluding section below a discussion of the possible consequences of our results for the general programme of 
 developing PM gravity to higher orders in $G$.

\section{Gravitational waveform at $O(G^2)$.} \label{sec2}
  
  The radiative part of the GR gravitational field  is asymptotically measured by the
  waveform, $f_{ij}(u, \theta, \phi)$  (where $u\equiv t-r$), defined as the transverse-traceless projection of the 
  $O(1/r)$ part of the metric\footnote{We use a mostly plus signature, and generally use $c=1$.}
   $g_{\mu \nu}= \eta_{\mu\nu} + h_{\mu\nu}$ in a suitable Bondi-Sachs-type coordinate system:
  \be
  h_{ij}^{\rm TT}= \frac{f_{ij}(t-r, \theta, \phi)}{r} + O\left( \frac1{r^2} \right) .
  \ee
 The outoing wavefield $ h_{ij}^{\rm TT}$ carries away both energy-momentum, $P^{\rm rad}_{\mu}$, 
 and angular momentum $J^{\rm rad}_{\mu \nu}$. There are subtleties in the definition of $J^{\rm rad}_{\mu \nu}$
 linked to nonlinear memory effects, but they do not affect our $O(G^2)$ computation below.
 Indeed, one of the crucial elements underlying our computation is that, while 
 $P^{\rm rad}_{\mu}= O(G^3)$, the radiated angular momentum is of lower PM order, namely:  $J^{\rm rad}_{\mu \nu}= O(G^2)$.
Expressions for the radiated angular-momentum\footnote{We work with the total radiated angular momentum, integrated
both over angles and (retarded) time, and evaluated in the center of mass (c.m.) frame.}  
in terms of the asymptotic waveform $f_{ij}(u, \theta, \phi)$ have been given
in Refs. \cite{Peters:1964zz,DeWitt:2011nnj,Thorne:1980ru,Bonga:2018gzr}
(see also \cite{Blanchet:2018yqa} for $J^{\rm rad}_{0 i}$)
\be \label{Jrad}
J^{\rm rad}_{k}=\frac{\epsilon_{kij}}{16\pi G} \int\!  du \, d\Omega \left[f_{ia} \partial_u  f_{ja}- \frac12 x^i \partial_j f_{ab}  \partial_u  f_{ab} \right].
\ee
Note that the integrand of Eq. \eq{Jrad} is bilinear in $f_{ij}$ (or its angular derivatives) and in  $\dot f_{ij} \equiv \partial_u f_{ij}$.
By contrast, the radiated energy-momentum, $P^{\rm rad}_{\mu}$, is quadratic in  $\dot f_{ij}$, namely
 \be \label{Prad}
P_{\rm rad}^{\mu}=\frac{1}{32\pi G} \int\!  du \, d\Omega \left[ \partial_u  f_{ab}  \partial_u  f_{ab} \right]n^{\mu} ,
\ee
where $n^{\mu}=(1, x^i/r)$.
The PM perturbation theory of gravitationally  interacting point masses \cite{Kovacs:1977uw,Kovacs:1978eu,Bel:1981be,Westpfahl:1985}
has shown that the waveform has an expansion
in powers of $G$ of the form
\be
f_{ij}(u, \theta, \phi)= G f^{(1)}_{ij}( \theta, \phi)+  G^2 f^{(2)}_{ij}(u, \theta, \phi)+ O(G^3) ,
\ee
where the 1PM ($O(G)$) contribution is independent\footnote{This is the way $f^{(1)}_{ij}$ is defined in 
Refs. \cite{Kovacs:1977uw,Kovacs:1978eu}.
Alternatively, in the definition of $h^{(1)}_{\mu \nu}$ in Ref. \cite{Bel:1981be}, the corresponding waveform $f^{(1)}_{ij}$ would be a function of $u$ that
varies on a slow, $G$-dependent  time scale determined by the gravitational interaction.} of the retarded time $u$, so that
\be
\dot f_{ij}= G^2 \dot f^{(2)}_{ij}  + O(G^3) ,
\ee
is of order $ O(G^2)$. This simple fact immediately shows, in view of Eqs. \eq{Jrad}, \eq{Prad}, that  $J^{\rm rad}_{\mu \nu}= O(G^2)$,
while $P^{\rm rad}_{\mu}= O(G^3)$.

The static nature of  $f^{(1)}_{ij}$ allows one to perform the time integration in Eq. \eq{Jrad} and to express the radiated angular momentum
as 
\be \label{Jrad2}
J^{\rm rad}_{k}=\frac{\epsilon_{kij}}{16\pi } \int\!   d\Omega \left[f_{ia}^{(1)} \Delta f_{ja}- \frac12 x^i \partial_j f_{ab}^{(1)}  \Delta  f_{ab} \right] ,
\ee
where 
\be
\Delta f_{ij}(\theta,\phi) \equiv \int_{- \infty}^{+ \infty} du \partial_u  f_{ij}= \left[  f_{ij} \right]_{u=- \infty}^{u=+\infty}\,,
\ee
is the {\it gravitational wave memory}, i.e., the global change in the 
waveform. To compute the waveform memory, $\Delta f_{ij}(\theta,\phi)$,  it is better to use the form of the PM expansion
employed, e.g., in Ref. \cite{Bel:1981be}, where the gothic-metric perturbation,  
${\bar h}^{\mu \nu} \equiv -( {\mathfrak g}^{\mu \nu} - \eta^{\mu \nu})$,
i.e., ${\bar h}^{\mu \nu}=\eta^{\mu \mu'}\eta^{\nu \nu'}{ h}_{\mu' \nu'}- \frac12 h \eta^{\mu \nu}+ O(h^2)$, 
is decomposed as ${\bar h}^{\mu \nu}= {\bar h}^{\mu \nu}_{\rm lin} + {\bar h}^{\mu \nu}_{\rm nonlin}$.
Here the linear piece, ${\bar h}^{\mu \nu}_{\rm lin}$, is the one generated by $ T^{\mu \nu}$, i.e., 
${\bar h}^{\mu \nu}_{\rm lin} = - 16\pi G\Box^{-1}_{\rm ret}({ T}^{\mu \nu}) $,
while ${\bar h}^{\mu \nu}_{\rm nonlin}$ is sourced by the nonlinear contributions ${ T}^{\mu \nu}_{\rm nonlin}= \partial^2 h h+   \partial^2 h h h+ \cdots$ appearing on the right-hand side of 
the PM-expanded Einstein's equations (in harmonic coordinates). The linear contribution generated by a system of pointlike bodies reads
\be \label{hlin}
{\bar h}^{\mu \nu}_{\rm lin}(x)= \sum_A  4 Gm_A \left( \frac{u_A^\mu  u_A^\nu}{r_A}\right)_{\rm ret}\,.
\ee
Here: the index $A$ labels the gravitationally interacting bodies ($A=1,2$ for our present case); $u_A^\mu= d z_A^\mu/d\tau_A$ denotes the 
(Minkowski-normalized) four velocities;  $r_A\equiv -  \eta_{\mu \nu} (x^\mu-z_A^{\mu}) u_A^\nu$; and the subscript ``ret" indicates
that $z_A$ is the retarded foot of the field point $x$ on the Ath worldline (such that $\eta_{\mu \nu} (x^\mu-z_A^{\mu})  (x^\nu-z_A^{\nu})=0$ and $x^0-z_A^{0}>0$) . In addition, the worldlines used in the definition Eq. \eq{hlin} are taken to be the exact worldlines,
or, at least,  1PM-accurate worldlines, curved by the linearized gravitational interaction. 

When using such a decomposition, it is physically clear that the nonlinear memory $\Delta f_{ij}^{\rm nonlin}$ induced by the
nonlinear effective source, i.e., by the  splash of gravitational-wave energy emitted by the collision, will be of order 
$G \times P^{\rm rad}_{\mu}$, and therefore $O(G^4)$ (as was explicitly shown in Ref. \cite{Wiseman:1991ss}).
We therefore conclude that the waveform memory $\Delta f_{ij}$ to be used in Eq. \eq{Jrad2} is given with sufficient accuracy
by the simple linear formula
\be \label{deltaf}
\Delta f_{ij}(\n)= 4 G   \left[  \frac{\left(p_A^i p_A^j\right)^{\rm TT}}{E_A- \n \cdot \p_A} \right]_{- \infty}^{+\infty}+O(G^4) .
\ee
Here: $\n = n^i(\theta, \phi)$ is the unit vector parametrizing the direction of gravitational-wave emission; $p_A^\mu=(E_A,  \p_A)$,
with $E_A= \sqrt{m_A^2+ \p_A^2}$, is the four-momentum of the Ath particle in the incoming (${- \infty}$)
or outgoing  (${+ \infty}$) state (with  $\left[f \right]_{- \infty}^{+\infty}\equiv f(+\infty) -   f(-\infty)$); and $f_{ij}^{\rm TT}$
denotes (as usual) the transverse-traceless projection of a (symmetric) three-dimensional tensor $f_{ij}$ in the two-plane
orthogonal to  $\n = n^i(\theta, \phi)$, i.e., the two-plane tangent to the sphere at infinity.

Note that it is crucial to insert in the expression Eq.~\eq{deltaf} values for  $p_A^\mu(\pm\infty)$ that take into account
the effect of the gravitational scattering. For computing the $O(G^2)$ angular-momentum loss, it is enough to use the $O(G)$
(1PM-accurate) gravitational deflection, i.e., (for $A=1,2$)
\be \label{deltap}
\Delta p_1^\mu=-\Delta p_2^\mu= - \beta \frac{b^\mu}{b} + O(G^2),
\ee
where we defined
\be \label{beta}
\beta \equiv \frac{2 (2 \g^2-1)}{\sqrt{\g^2-1}} \frac{Gm_1 m_2}{b}\,.
\ee
Here, 
\be
\g \equiv - (u_1 \cdot u_2)_{- \infty} = -  \left[\frac{p_1 \cdot p_2}{m_1 m_2}\right]_{- \infty},
\ee
denotes the  Lorentz factor between the two incoming worldlines, while
$b^\mu $ (of magnitude $b$) denotes the vectorial impact parameter, i.e., the value of the
four-vector $z_{1 (0)}^{\mu}(\tau_1) - z_{2 (0)}^{\mu}(\tau_2)$ {\it orthogonally} connecting the incoming (unperturbed)
worldlines. [$b^\mu $ is orthogonal both to $p_1^\mu(-\infty)$ and to $p_2^\mu(-\infty)$; see \cite{Bini:2018ywr} for details.]

\section{Waveform and waveform-memory in the center-of-mass frame}

We are interested in computing the change of angular momentum of the binary system in the (incoming) center-of-mass (c.m.)
frame, i.e., with $ \p_1(-\infty)=-\p_2(-\infty) \equiv \bP$, and $E_A(-\infty)= \sqrt{m_A^2+ \bP^2}$.
We recall that $P_{\rm rad}^{\mu}=O(G^3)$, so that 
we have also $E_A(+\infty)= E_A(-\infty)+ O(G^3)$, and 
$\bP' = \p_1(+\infty)+ O(G^3)=-\p_2(+\infty) + O(G^3)$, where $\bP'$ is the deflected value of $\bP$,
obtained by rotating it by the c.m. deflection angle $\chi$.
The magnitude of the incoming c.m. angular momentum is $J(- \infty)= b P$, where $P \equiv |\bP|$.

Using Eq. \eq{hlin} the waveform is given, to $O(G)$ accuracy, by the time-independent TT tensor
\be \label{f1PM}
 f_{ij}(\n)= 4 G  \left[  \frac{1}{E_2+ \n \cdot \bP}+ \frac{1}{E_1- \n \cdot \bP} \right]\left(P^i P^j\right)^{\rm TT}  +O(G^2). 
\ee
On the other hand, the waveform memory is obtained by inserting the changes \eq{deltap} in the expression \eq{deltaf}, with the result
\bea
\Delta f_{ij}(\n) \! \! \! \!&&= -\beta \frac{4G}{b} \left[ \frac{1}{E_2+ \n \cdot \bP} + \frac{1}{ E_1- \n \cdot \bP}  \right] \left(P^i b^j+ P^j b^i  \right)^{\rm TT} \nonumber\\
&& + \beta \frac{4G}{b} \left[ \frac{\n \cdot \bb}{(E_2+ \n \cdot \bP)^2}- \frac{\n \cdot \bb}{(E_1- \n \cdot \bP)^2} \right] \left(P^i P^j \right)^{\rm TT}.\nonumber\\
\eea

\section{Radiated angular momentum}

In order to explicitly compute the angular integral \eq{Jrad} giving the radiated angular momentum, it is convenient to choose
a system of polar coordinates adapted to the angular dependence of $ f_{ij}(\n)$ and $\Delta f_{ij}(\n) $. We choose 
$\ve_z$ in the direction of $\bP$ (i.e., $\bP = + P \ve_z$), and $\ve_y$ in the direction of $\bb$ (i.e., $\bb = + b \ve_y$).
We then define polar coordinates $\theta, \phi$ in the usual way i.e., $\n= (\sin \theta \cos \phi, \sin \theta \sin \phi , \cos \theta)$, 
as well as a standard orthonormal two-frame tangent to the sphere, namely 
$\ve_{\theta}= (\cos \theta \cos \phi, \cos \theta \sin \phi , -\sin \theta)$
and  $\ve_{\phi}= (- \sin \phi, \cos  \phi , 0)$. The two independent components of the waveform with respect to the frame 
$\ve_{\theta}$, $\ve_{\phi}$ are then $ f_+ \equiv \frac12 (f_{\theta \theta} - f_{\phi \phi})$ and  $ f_{\times} \equiv \frac12 (f_{\theta \phi} + f_{\phi \theta})$.

Inserting $\n \cdot \bP = P \cos \theta$, and  introducing the short-hand notations,
\bea
A &\equiv&   \frac{P}{E_2 + P \cos \theta} + \frac{P}{E_1 - P \cos \theta}, \nonumber\\
B  &\equiv&   \frac{P^2}{(E_2 + P \cos \theta)^2} - \frac{P^2}{(E_1 - P \cos \theta)^2},
\eea
the $+$ and $\times$ components of Eqs. \eq{deltaf} and \eq{f1PM} read
\be \label{f+x}
f_+=2 \,G P A  \sin^2 \theta +O(G^2)\; ; \; f_{\times}= O(G^2) \,,
\ee
and
\bea \label{deltaf+x}
\Delta f_+ &=& 4G \beta \left[ A \sin \theta \cos \theta \sin \phi + \frac{B}{2} \sin^3 \theta \sin \phi \right], \nonumber\\
\Delta f_{\times} &=& 4G \beta  A \sin \theta \cos \phi .
\eea
Re-expressing  Eq. \eq{Jrad2} in terms of $f_{+, \times}$ yields,  for the relevant
component $J_x^{\rm rad} = J_{y z}^{\rm rad}$ in the direction of the initial angular momentum  $J_{y z}^{(0)}= b P$, the expression \cite{Bonga:2018gzr}
\bea \label{Jradf+x}
J_{y z}^{\rm rad}&=&\int \frac{ d\Omega}{16 \pi G \sin \theta}\left[ ({\cal D} f_+ - 2 \cos \phi f_{\times}) \Delta f_+ \right. \nonumber\\
&&  \left. + \, ({\cal D} f_{\times} + 2 \cos \phi f_+) \Delta f_+\right]\,,
\eea
where $ {\cal D} \equiv \sin \theta \sin \phi \partial_{\theta} + \cos \theta \cos \phi \partial_{\phi}$.

Inserting Eqs. \eq{f+x}, \eq{deltaf+x} in Eq. \eq{Jradf+x} yields an explicit integral which can be performed without
difficulty. [The integral over $\phi$ is elementary, while the integral over $\theta$ becomes elementary when rewritten
as an integral over $c \equiv \cos \theta$.] Though intermediate results depend on the c.m. velocities of the
two bodies, i.e., $v_1 = P/E_1$ and $v_2=P/E_2$,  the final result only depends on the (relativistic) relative velocity $v$
between the two bodies, namely
\be
v \equiv \frac{v_1+v_2}{1+ v_1 v_2} = \sqrt{1- \frac1{\g^2}} \,,
\ee
and takes the relatively simple form
\be
J_{y z}^{\rm rad} = \frac{2 (2 \g^2-1)}{\sqrt{\g^2-1}} \frac{G^2 m_1 m_2 P}{b} {\cI}(v) + O(G^3)\,.
\ee 
Here ${\cI}(v)$ denotes the following function
\be
{\cI}(v) = -\frac{16}{3}+ \frac{2}{v^2}+ \frac{2(3 v^2-1)}{v^3} \cA(v)\,,
\ee
where $\cA$ is a  short-hand notation  for the arctanh function, i.e.,
\be \label{cA}
\cA(v) \equiv {\rm arctanh}(v) = \frac 12 \ln \frac{1+v}{1-v} =2 \,{\rm arcsinh} \sqrt{ \frac{\g-1}{2}}.
\ee
The last form relates $\cA(v) $ to the (crucial) arcsinh function entering the 3PM results of Refs. \cite{Bern:2019nnu,Bern:2019crd}.
[The latter works use the notation $\sigma$ for the Lorentz factor here denoted $\g = - (u_1 \cdot u_2)_{- \infty}$.]

The ratio between the radiated angular momentum, and the incoming mechanical angular momentum of the binary system, 
takes the following simple form
\be \label{JradbyJ}
\frac{J^{\rm rad}}{J}=\frac{J_{y z}^{\rm rad} }{b P} = \frac{2 (2 \g^2-1)}{\sqrt{\g^2-1}} \frac{G^2 m_1 m_2 }{b^2} {\cI}(v) + O(G^3)\,.
\ee
The PN (slow velocity) expansion of Eq. \eq{JradbyJ} begins as
\be
\frac{J^{\rm rad}}{J}= \frac{G^2 m_1 m_2 }{c^5 b^2} \left( \frac{16}{5}  v + \frac{232}{35} \frac{v^3}{c^2} + \frac{2146}{315} \frac{v^5 }{c^4}+ \cdots\right),
\ee
where we added the (dimensionally determined) powers of $\frac1c$ to emphasize that this is a 2.5PN ($O(G^2/c^5)$) effect.
The leading-order term in $\frac{J^{\rm rad}}{J}$ agrees with the result of Ref. \cite{Damour:1981bh} (where it was directly derived
from the $G^2$-accurate retarded equations of motion of the binary system).  We have also checked that the next-to-leading-order term
is compatible with the fractionally 1PN-accurate computation \cite{Junker1992} of the angular momentum radiated to gravitational waves during
an hyperbolic encounter. Let us note that the function $\cI(v)$ is positive, and monotonically growing, in the interval $0<v<1$.

\section{Radiation-reaction contribution to the scattering angle.}

Having in hand the leading-order radiative loss of angular momentum, we can deduce from it the
corresponding radiative contribution to the scattering angle. Indeed, Ref. \cite{Bini:2012ji} has derived a
general formula, namely Eq. (5.74) there\footnote{As indicated around Eq. (5.98) of  Ref. \cite{Bini:2012ji} this formula has
a general validity.},
yielding the radiation-reaction contribution to the scattering angle (considered as a
first-order correction to the conservative-dynamics value of the scattering angle). Namely,
\be \label{chiradgen}
\chi^{\rm rad}(E, J)=-\frac12 \frac{\partial \chi^{\rm cons}}{\partial E} E^{\rm rad} -\frac12 \frac{\partial \chi^{\rm cons}}{\partial J} J^{\rm rad} \,.
\ee
Note here the factor $\frac12$ and the negative sign, because $E^{\rm rad}$ and $J^{\rm rad}$ denote the (positive) energy and
angular momentum radiated away. [$E$ and $J$ are both measured in the c.m. frame.]

As $E^{\rm rad}= O(G^3)$, while $J^{\rm rad}= O(G^2)$, the leading-PM-order contribution to $\chi^{\rm rad}(E, J)$ will be the one
induced by the angular-momentum loss. In addition, the leading-PM-order contribution to $\chi^{\rm cons}(E,J)$ is $\propto G J^{-1}$,
so that the leading-PM-order contribution to $\chi^{\rm rad}(E, J)$ is {\it positive}, of order $ O(G^3)$, and given by
\be
\chi^{\rm rad}(E, J) = + \frac12 \chi^{\rm cons}_{\rm LO} \frac{J^{\rm rad}}{J}\,.
\ee
Inserting our result \eq{JradbyJ}, and working (as is often convenient) with the half-scattering angle,
\be \label{chiPMcons}
\frac12 \chi^{\rm cons}(E,J)= \frac{\chi_1(\g)}{j}+ \frac{\chi_2(\g)}{j^2}+ \frac{\chi_3(\g)}{j^3}+ \cdots
\ee
where $j \equiv J/(G m_1 m_2)$ and
\be
\chi_1(\g)=\frac{2 \, \g^2-1 }{\sqrt{\g^2-1}}\,,
\ee
we get the explicit result
\be \label{chirad2}
\frac12 \chi^{\rm rad}= + \frac{(2 \, \g^2-1)^2 }{\g^2-1} \cI(v) \frac{G^2 m_1 m_2}{b^2 j}+ O(G^4)\,.
\ee
Using the relation \cite{Damour:2017zjx}
\be
\frac{GE}{b} \equiv \frac{GM h(\g,\nu)}{b} = \frac{\sqrt{\g^2-1}}{j}\,,
\ee
 where $h(\g,\nu)\equiv \frac{E}{M}=\sqrt{1+ 2\nu (\g-1)}$, we can rewrite \eq{chirad2} either in terms
 of $b$ or of $j$. Its expression in terms of $j$ reads
 \be \label{chirad3}
\frac12 \chi^{\rm rad}(\g, j,\nu)= + \frac{\nu }{ h^2(\g,\nu) j^3} (2 \, \g^2-1)^2  \cI(v)+ O(G^4)\,,
\ee
where $\nu\equiv \frac{m_1m_2}{(m_1+m_2)^2}$ denotes the symmetric mass ratio.
 
 \section{Radiation-reaction corrected scattering angle}
 
 The $O(G^3)$ contribution, $\chi_3/j^3$, to the (half) {\it conservative} scattering angle (see Eq. \eq{chiPMcons})
 has been computed by Bern et al.  \cite{Bern:2019nnu,Bern:2019crd}
 (see also \cite{Kalin:2020fhe}) with the result
\be \label{chi3vsbarC}
\chi_3^{\rm cons}(\g,\nu)= \chi_3^{\rm Schw}(\g)    - \frac{ 2 \,\nu \, p_{\infty} }{h^2(\g,\nu)}  {\overline C}^{\rm cons}(\g)\,,
\ee
where (denoting $ \pinf \equiv \sqrt{\g^2-1}= \frac{v}{\sqrt{1-v^2}}$)
\be \label{chi3Schw}
\chi_3^{\rm Schw}(\g)=\frac{64\, \pinf^6 + 72 \,\pinf^4 + 12\, \pinf^2 -1}{3 \, \pinf^3},
\ee
and
\bea\label{bCB}
{\overline C}^{\rm cons}(\g) &=& \frac{2}{3} \g  (14 \g^2+25) \nonumber\\
&+& 2 (4 \g^4 - 12 \g^2 -3) \frac{ \cA(v)}{\sqrt{\g^2-1}}\,.
\eea
In the latter expression we have replaced the function ${\rm arcsinh} \sqrt{ \frac{\g-1}{2}}$ used in Refs. \cite{Bern:2019nnu,Bern:2019crd}
by the (half) arctanh function $\frac 12 \cA(v)$ (see Eq. \eq{cA}).

The physical, $O(G^3)$-accurate, total scattering angle, i.e., the sum of the conservative contribution and of the leading-PM-order radiation-reaction
correction \eq{chirad3}, then reads
\be \label{chiPMtot}
\frac12 \chi^{\rm tot}(E,J)= \frac{\chi_1(\g)}{j}+ \frac{\chi_2(\g)}{j^2}+ \frac{\chi_3^{\rm tot}(\g)}{j^3}+ O(G^4)\,,
\ee
where
\be
\chi_3^{\rm tot}(\g, \nu)= \chi_3^{\rm cons}(\g,\nu)+ \chi_3^{\rm rad}(\g,\nu)\,,
\ee
with
\be
\chi_3^{\rm rad}(\g,\nu)= + \frac{\nu }{ h^2(\g,\nu) } (2 \, \g^2-1)^2  \cI(v)\,,
\ee
or more explicitly
\bea
\chi_3^{\rm rad}(\g,\nu) &=&+ \frac{ 2\nu }{ h^2(\g,\nu) } \frac{(2 \, \g^2-1)^2 }{3 (\g^2-1)} \times \nonumber\\
&& \times \left(8-5\g^2 + (6 \g^2-9) \frac{\cA(v)}{v} \right)\,.
\eea
Let us note that the mass-ratio dependence of $\chi_3^{\rm rad}(\g,\nu)$ satisfies the general rule pointed out in Ref. \cite{Damour:2019lcq},
namely $h^2(\g,\nu)  \chi_3(\g,\nu)$ is a linear function of $\nu$. In other words, we can write $\chi_3^{\rm rad}(\g,\nu)$ in the same
form as $\chi_3^{\rm cons}(\g,\nu)- \chi_3^{\rm Schw}(\g)  $, namely
\be
\chi_3^{\rm rad}(\g,\nu)=  - \frac{ 2 \,\nu \, p_{\infty} }{h^2(\g,\nu)}  {\overline C}^{\rm rad}(\g)\,,
\ee
where
\bea
{\overline C}^{\rm rad}(\g) &=& - \frac{(2 \, \g^2-1)^2 }{ 2 \sqrt{\g^2-1}} \cI(v) \nonumber\\
&=&  - \frac{(2 \, \g^2-1)^2 }{  \sqrt{\g^2-1}} \left( -\frac{8}{3}+ \frac{1}{v^2}+ \frac{3 v^2-1}{v^3} \cA(v) \right). \nonumber\\
\eea

\section{Low-energy and high-energy limits of  the radiation-reaction corrected scattering angle}

 \subsection{Low-velocity limit}

 In the low-kinetic-energy (or low-velocity) limit $ v \to 0$ (or $\g \to 1$), the $v$-expansion of $ {\overline C}^{\rm cons}(\g)$ reads
 \be
 {\overline C}^{\rm cons}(\g)= 4 + 18 v^2 + \frac{271}{10} v^4 + \frac{4999}{140} v^6 +\frac{440273}{10080} v^8 + \cdots
 \ee
 The first term of this  expansion corresponds to the 2PN level. The last term we explicitly wrote belongs to the 6PN level. All
 those terms have been explicitly checked by PN-based computations \cite{Bini:2019nra,Blumlein:2020znm,Bini:2020wpo}.
 
 The corresponding PN expansion of the complementary radiation-reaction contribution $ {\overline C}^{\rm rad}(\g)$ reads
\be
 {\overline C}^{\rm rad}(\g)= - \frac{4}{5} v  - \frac{114}{35} v^3  - \frac{4169}{630}  v^5 - \frac{138451}{13860} v^7 - \cdots
 \ee 
 Here, the first term of this expansion corresponds to the 2.5PN level (which is indeed the leading level for radiation-reaction in GR).
 The corresponding leading-order (both in the PN and the PM senses),  $O(G^3/c^5)$, contribution to the scattering angle, namely
 \be
\frac12  \chi^{\rm rad} = +\frac{8 G^3  }{5 c^5}  \frac{m_1^3 m_2^3}{J^3} \nu v^2 + \cdots
 \ee
 agrees with the large-eccentricity limit of Eq. (5.116) in Ref. \cite{Bini:2012ji}.
 
 The radiation-reaction character of ${\overline C}^{\rm rad}(\g)$
 shows up in the fact that it involves only odd powers of the relative velocity $v$. In other words, ${\overline C}^{\rm rad}(\g)$
 is an odd function of $v$, while $ {\overline C}^{\rm cons}(\g)$ is an even function of $v$ (as is easily checked on their exact
 expressions).

 \subsection{High-energy (or massless) limit}
 
 Let us now consider the high-energy (HE) limit ($\g \to + \infty$ or $v \to 1$) of the scattering angle. This limit is taken
 together with the massless limit, $m_1 \to 0$, $m_2 \to 0$, keeping fixed the c.m. linear momentum $P$,
 so that $ \sqrt{s} = E = \sqrt{P^2+ m_1^2}+  \sqrt{P^2+ m_2^2} \to 2 P$. A natural PM expansion parameter in this
 limit is then
 \be \label{defalpha}
\alpha \equiv \frac{\g}{j} = \frac{G}{2} \frac{s - m_1^2  -m_2^2 }{J}\,.
\ee
In the HE limit we can write
 \be
 \alpha \overset{\rm HE}{=} \frac{G E}{b}\,.
 \ee
 The HE limit of the 3PM-accurate conservative scattering angle Eq. \eq{chiPMcons} (with \eq{chi3vsbarC}) reads
 \bea \label{chiconsHE}
 \frac12 \chi^{\rm cons} \overset{\rm HE}{=} && 2 \alpha + \left(\frac{64}{3}  -\frac{28}{3} - 8 \ln(2 \g) \right) \alpha^3 + O(G^4), \nonumber\\
 =&&  2 \alpha + \left(12 - 8 \ln(2 \g) \right) \alpha^3 + O(G^4),
 \eea
where the  $\ln(2 \g) $ term introduces a logarithmic divergence of $\frac12 \chi^{\rm cons}(\alpha)$ in the HE limit.
On the first line of  Eq. \eq{chiconsHE}, we have indicated the (positive) contribution from the 3PM-level Schwarzschild term 
$\chi_3^{\rm Schw}(\g)$, and the (negative) contribution from the $\nu$-dependent, last term in Eq. \eq{chi3vsbarC}.

On the other hand, the HE limit of the radiative correction to $\frac12 \chi$ reads
\be\label{chiradHE}
\frac12 \chi^{\rm rad} \overset{\rm HE}{=}  + \left( -\frac{20}{3} + 8 \ln(2 \g) \right) \alpha^3 + O(G^4),
\ee
As we see, the HE limit of $\frac12 \chi^{\rm rad} $ contains exactly the opposite of the logarithmic divergence 
contained in $ \frac12 \chi^{\rm cons}$. The radiation-reaction corrected scattering angle is therefore finite
in the HE limit, and actually equal to
\be \label{chitotHE}
 \frac12 \chi^{\rm tot} \overset{\rm HE}{=} 2 \alpha  + \frac{16}{3} \alpha^3 + O(G^4).
\ee
Remarkably, this HE limit of the radiation-reaction corrected scattering angle agrees with
the ACV eikonal-approximation two-loop result  \cite{Amati:1990xe},
namely (see \cite{Ciafaloni:2014esa,Bern:2020gjj})
\be \label{chiACV}
 \frac12 \chi^{\rm eikonal} \overset{\rm HE}{=}  2 \alpha + \frac{16}{3} \alpha^3 + O(\alpha^5).
\ee
The eikonal result \eq{chiACV} was initially derived in GR. Its validity in GR
was recently confirmed \cite{Bern:2020gjj,DiVecchia:2020ymx}. In addition,
it was proven to hold in all supergravity theories \cite{DiVecchia:2019kta,Bern:2020gjj,DiVecchia:2020ymx}.

The (finiteness and) equality, at the $G^3$ level, between the HE limit of
the {\it classical} radiation-reaction corrected scattering angle 
\eq{chitotHE}, and the {\it quantum-derived}, two-loop eikonal massless scattering angle \eq{chiACV} is the
main result of the present work.  Various aspects of this result are discussed next.

\section{Discussion}

Let us summarize the context, and meaning, of our derivation. We worked within a purely {\it classical} GR framework, 
and our derivation of the radiative correction to the scattering angle used a  post-Minkowskian (PM) perturbative
approach. By contrast, we are not aware of a fully classical, and fully PM, derivation of the so-called {\it conservative}
3PM-accurate scattering angle \eq{chi3vsbarC}. The 1PM, $O(G)$,  and 2PM, $O(G^2)$, scattering angles have
been computed within classical PM frameworks in Refs. \cite{Portilla:1980uz,Damour:2016gwp} (1PM)
and Refs. \cite{Westpfahl:1985,Bini:2018ywr} (2PM).  The first derivation \cite{Bern:2019nnu,Bern:2019crd} 
of the conservative 3PM scattering angle was performed by using a mix of various tools: a selection of 
quantum scattering amplitude integrands, and an evaluation of the corresponding integrals by expanding
the integrand in the so-called {\it potential region}. The rederivation of Ref. \cite{Kalin:2020mvi} cannot either
be considered as a purely classical PM derivation because the propagator of the graviton used in the latter computation
is the Feynman one, namely $ \propto (k^2 - i 0)^{-1}$. 
Let us recall in this respect that the correct, classical 
 conservative action for PM gravity \cite{Damour:1995kt} must be classically defined, 
 \`a la  Fokker-Wheeler-Feynman \cite{Fokker1929,Wheeler:1949hn}, by using the {\it time-symmetric}
 graviton propagator, whose Fourier-space kernel is
 \be \label {Gsym}
 G_{\alpha \beta ;\alpha' \beta'}(k)= (\eta_{\alpha \alpha'} \eta_{\beta \beta'} -\frac12 \eta_{\alpha \beta} \eta_{\alpha' \beta'}) {\rm PP}\frac1{k^2},
 \ee
 where PP denotes the principal value. By contrast, the Feynman propagator involves 
 \be \label{GF}
 \frac1{k^2 - i 0}= {\rm PP}\frac1{k^2} + i \pi \delta(k^2)\,,
 \ee
 while the retarded propagator would involve
 \be \label{Gret}
 \frac1{k^2 - {\rm sign}(k^0) i 0}= {\rm PP}\frac1{k^2} + i \pi {\rm sign}(k^0) \delta(k^2).
 \ee
 The differences between the coefficients of the $\delta(k^2)$ term are crucial here, and start making a difference
 already at the 2PM, $O(G^2)$, level. Indeed, the 2PM-accurate {\it retarded} equations of motion derived in
 Refs. \cite{Westpfahl:1979gu,Bel:1981be} lead (as was explicitly shown in Ref. \cite{Damour:1981bh}) to an 
$O(G^2)$ loss of the {\it mechanical} angular momentum of the binary system, which was checked to balance the {\it radiated}
 angular momentum {\it at lowest PN order}. [We have used here the fact that this balance must formally hold 
 at the full (PN-exact) 2PM level, leaving a direct technical check to future work.]

At the 3PM level, the effects linked to using either a time-symmetric propagator, a Feynman one, or a retarded one
should be even more drastic. The first derivation \cite{Bern:2019nnu,Bern:2019crd} of the conservative 3PM
dynamics (as well as the rederivation of Ref. \cite{Kalin:2020mvi}) selected a graviton propagator by expanding the
graviton propagator in the so-called potential region, defined by the inequality $k^0 \ll | \vk |$. At face value,
it seems that the use of such an expansion, namely (with $\omega \equiv k^0$)
\be \label{Gpotential}
\frac1{k^2 + i \epsilon} = \frac1{\vk^2 - \omega^2  + i \epsilon} = \frac1{\vk^2} + \frac{\omega^2}{\vk^4} + \frac{\omega^4}{\vk^6}+ \cdots
\ee
is not only independent of the choice of contour around the poles in $\omega$ (i.e. of the choice of $ i \epsilon$),
but is equivalent to the PN-expansion of the $x$-space time-symmetric massless propagator
\be \label{GPN}
\Box^{-1}_{\rm sym}=(\Delta - \partial_0^2)^{-1}_{\rm sym}= \Delta^{-1} +  \partial_0^2\Delta^{-2}+ \partial_0^4\Delta^{-3}+\cdots
\ee
This suggests that computing the dynamics by expanding in the potential region is equivalent to using the PN-expanded
time-symmetric graviton propagator. As genuine nonlocal-in-time effects enter only at the 4PN and 4PM ($O(G^4/c^8)$) level 
\cite{Blanchet:1987wq}, this indicates that the potential-region expansion should yield the PN-expanded version
of the conservative dynamics up to the accuracy $O(G^3)$, but will encounter subtle nonlocal-in-time effects starting
at the $O(G^4)$ level. For our present purposes, it confirms that the current potential-region derivations of the 3PM scattering 
angle \eq{chi3vsbarC} do capture the classical (Fokker-Wheeler-Feynman-type) time-symmetric conservative dynamics.
It would, however, be instructive to give both an {\it ab initio} classical derivation of the 3PM conservative dynamics,
as well as an {\it ab initio} classical derivation of the retarded 3PM dynamics, so as to directly check that the radiation-reaction
corrected scattering angle \eq{chiPMtot} does directly follow from the retarded 3PM dynamics, without appealing (as we did)
to the balance between mechanical and radiated angular momentum.

As already mentioned in the Introduction, the motivation for the present work was Ref. \cite{DiVecchia:2020ymx} which
pointed out two different, but interrelated, facts: (i) analyticity, crossing properties, and impact-parameter-space
exponentiation of the HE quantum scattering amplitude allows one to relate the real part of the two-loop eikonal phase to its imaginary part;
the latter being then derived from the phase-space integral of a three-particle cut; and (ii) the explicit evaluation of  the four-point two-loop amplitude in  ${\cal N}=8$ supergravity using integration over the full soft-graviton region $k^0 \sim |\vk| \ll \sqrt{s}$. Their first approach
led to a rederivation of the ACV HE eikonal phase that emphasized its link with the {\it inelastic} tree-level 
amplitude describing the emission of a graviton in a certain HE double-Regge limit. Their second approach confirmed (in a particular
setting) that radiative corrections (soft region, rather than potential region) were crucial for getting a finite HE
eikonal phase. Our present work has completed, on the classical side, the quantum-based results of Ref. \cite{DiVecchia:2020ymx}
by showing in detail how classical radiative corrections to the classical conservative dynamics\footnote{In this sense, we do not
agree with the statement of Ref. \cite{DiVecchia:2020ymx} that ``the real part of  [the two-loop eikonal phase $\delta_2$] 
captures the conservative dynamics." Our calculation clearly shows that the $J$-derivative of the HE eikonal phase 
is the full, radiation-reaction-corrected scattering angle.}  resolve the puzzle
of the logarithmic divergence in the HE (or massless limit) of the conservative 3PM scattering angle, by establishing,
by a purely classical computation, that the ACV eikonal scattering does agree with the massless limit of the 3PM
scattering of two classical masses.

Let us note that, independently of the issue of the high-energy behavior, our results show that even the 2PM ($O(G^2)$) 
dynamics cannot be considered as being naturally conservative (as had been assumed so far because of the $O(G^3)$ 
order of the radiative energy-momentum loss $P_{\mu}^{\rm rad}$). Indeed,  the 
 2PM-accurate effective one-body Hamiltonian description \cite{Damour:2017zjx} (or the equivalent \cite{Damour:2019lcq}
2PM effective-field-theory formulation \cite{Cheung:2018wkq}) involves the total c.m. angular momentum $J$
of the binary system, and crucially assumes that $J$ is conserved. Our explicit PM result \eq{JradbyJ} above shows that
that this is not true for the physical (retarded) PM interaction, but only applies to the  time-symmetric
2PM dynamics. Another consequence
 of our result is that, contrary to what had been hitherto assumed, the classical ACV
 scattering angle should not be thought as being a conservative quantity, but rather as a strongly-radiation-reacted quantity.
 The ACV scattering angle should not therefore be transcribed into some high-energy Hamiltonian (as was done, e.g., in \cite{Damour:2017zjx}).
 At the formal level, one could use the dictionary given in Refs. \cite{Damour:2017zjx,Damour:2019lcq,Bini:2020wpo} to transcribe,
 for any finite values of the masses and of $\g$, the 3PM-level radiation-corrected total scattering angle $\chi^{\rm tot}_3$ into
 a corresponding value for the 3PM-level effective one-body potential (either in the $Q$ Finsler-like form or the $W$ potential one). The resulting
 potential would be of the form $\hat Q^E= q_2(\g,\nu) (GM/R)^2 + q_3^{\rm tot}(\g,\nu) (GM/R)^3 +O(G^4)$,
 where $q_2(\g,\nu)$ is given in Eq. (5.12) of \cite{Damour:2017zjx}, and where
 $q_3^{\rm tot}(\g,\nu) = q_3^{\rm cons}(\g,\nu) + q_3^{\rm rad}(\g,\nu) $, where $ q_3^{\rm cons}(\g,\nu)$ is the 
 3PM-level conservative effective one-body potential (see \cite{Antonelli:2019ytb,Damour:2019lcq}),
 and where the radiative correction would explicitly read
 \bea
 q_3^{\rm rad}(\g,\nu) &=& \frac{2 \nu}{h^2(\g,\nu)} {\overline C}^{\rm rad}(\g)\nonumber\\
 &=& - \frac{2 \nu}{1 + 2\nu(\g-1)} \frac{(2\g^2-1)^2}{\sqrt{\g^2-1}} \times\nonumber\\
 && \times \left( -\frac{8}{3}+ \frac{1}{v^2}+ \frac{3 v^2-1}{v^3} \cA(v) \right).
 \eea
 The latter radiative contribution to the $Q$ potential would start at the 2.5PN order (namely 
 $ q_3^{\rm rad}= - \frac85 \nu  (GM/R)^3 (v+ O(v^3))$) and would be odd in the relative velocity $v$. 
 It would be (somewhat analogously to the use of an optical potential) a way of encoding the effect of radiation-reaction on $\chi$ 
 in an effective potential.
 But it would miss the fact that the angular momentum of the system is not conserved at the $G^2$ level.
 For concreteness, let us recall that the leading-order {\it radiation-reaction force} (to be added to the conservative acceleration of
 the first body) is of order $ G^2/c^5$, and reads (see Eq. (5a) in \cite{Damour:1981bh}) 
 \be
 {\mathbf a}_1^{\rm rad}= - \frac45 \frac{G^2}{c^5} m_1 m_2 \frac{v_{12}^2}{r_{12}^3} \left[ {\mathbf v}_{12} - 3 ({\mathbf v}_{12}\cdot {\mathbf n}_{12}) {\mathbf n}_{12}\right]\,,
 \ee
where ${\mathbf v}_{12}= {\mathbf v}_{1} - {\mathbf v}_{2}$, and ${\mathbf n}_{12}= ({\mathbf x}_{1} - {\mathbf x}_{2})/r_{12}$.

The  time-symmetric dynamics of gravitationally interacting systems does admit a 
perturbatively defined Poincar\'e-invariant action \cite{Damour:1995kt}, with Noether-associated conservation laws for
energy-momentum and (relativistic) angular momentum (see, e.g.,  \cite{Wheeler:1949hn} for the electromagnetic case).
We recalled in the Introduction that in the 1980's it had been important to use a (physical, retarded) PM framework in order
to establish, without using any ill-defined PN-type nearzone expansion, that GR did predict nearzone radiation-reaction effects
in accord with the heuristic expectation of a global balance between the mechanical energy and angular momentum of the
binary system, and the energy and angular momentum radiated in gravitational waves. However, the classical GR community
stopped pursuing to higher orders the physical (retarded) PM approach because of the following combination of facts:
(i)  technical difficulties in explicitly evaluating retarded PM equations of motion at the  3PM, $O(G^3)$, level (which
could only be done by neglecting some terms explicitly involving the relative velocities between the two bodies \cite{Damour:1982ik});
(ii)  confidence built by the PM result in the validity of the nearzone-based PN expansion (whose breakdown was, moreover, 
shown to occur only at the $G^4/c^8$ level \cite{Blanchet:1987wq}); and (iii)   superior technical efficiency of the PN approximation for computing higher-order
effects \cite{Blanchet:1995ez,Jaranowski:1997ky,Blanchet:2000ub}. 
Instead of using a physical, retarded PM approach, it was found efficient to rely on a  dual approximation where the equations of motion of
binary systems are decomposed in a conservative part and a radiation-reaction one, both parts being computed by using 
PN methods (amplified by a PN-matched multipolar-post-Minkowskian approach, see, e.g., \cite{Blanchet:2013haa} for a review). 
[Such a dual approximation is notably used in the effective one-body approach \cite{Buonanno:1998gg,Buonanno:2000ef},
and has been recently pushed to very high orders \cite{Bini:2020nsb,Bini:2020hmy}.]

It is only recently that the PM approach (especially in its quantum-amplitude version) was rekindled with the assumption 
\cite{Cheung:2018wkq,Bern:2019nnu,Bern:2019crd}
that the combined use of quasi-classical-type approximations, and potential-region truncation, would  give an efficient (and scalable) way to
compute higher-orders in PM gravity. This enthusing programme is, however, now facing several types of difficulties highlighted by 
various recent results (notably Refs. \cite{Bini:2019nra}, \cite{DiVecchia:2020ymx} and the results presented here).
We have in mind here both the  subtleties linked to the differences between the various graviton propagators, as displayed
in Eqs. \eq{Gsym}, \eq{GF}, \eq{Gret}, and the breakdown of the potential-region propagator Eq. \eq{Gpotential} (or 
its $x$-space, PN-expanded, analog  Eq. \eq{Gpotential}) starting at the $G^4/c^8$ level, where time-nonlocality becomes
essential \cite{Blanchet:1987wq}.  Now that these subtlelties have been more clearly identified, we hope that improved ways
of tapping the deep knowledge of quantum perturbative gravity brought by many years of work (notably related to string theory)
for deriving classical observable quantities of direct importance for gravitational-wave physics will be explored. The dual,
conservative-plus-radiation-reaction, {\it PM-based} approach (instead of the usual {\it PN-based} one), exemplified by our
computation above, is {\it a priori} scalable, i.e., can be extended to higher PM orders.  We leave to future work an application 
of our method to the $O(G^4)$ level.

Let us finally note that  it is tempting (as suggested in Ref. \cite{DiVecchia:2020ymx})
to assume that Weinberg's quantum result \cite{Weinberg:1965nx} about
the absence of $\ln m_A$ divergences in the massless limit ($m_A \to 0$) of perturbative quantum-gravity amplitudes  
implies, at the classical level, that the physical, retarded (radiation-reaction corrected) scattering angle admits,
at each order  of the (classical) PM perturbative gravity expansion,  a finite massless limit, and therefore a well-defined HE limit.
 We wish, however, to recall that
Section VI D of Ref. \cite{Damour:2019lcq} has presented an argument (based on the results
of Refs. \cite{Gruzinov:2014moa,Ciafaloni:2015xsr}) showing how  radiative (tail) effects give a  {\it conservative} contribution 
to the scattering angle of order $G^4$ at finite $\g$, which becomes of order $G^3 \ln G$ in the HE limit $\g \to \infty$.



\end{document}